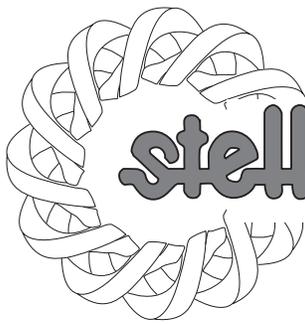

# stellarator news

## Report of the Office of Technology Assessment of the Congress of the United States

The following text is taken from the report of the Office of Technology Assessment of the Congress of the United States, titled *The Fusion Energy Program: The Role of TPX and Alternate Concepts*.

The Web version was dated January 31, 1995. The full report is available from the World-Wide Web at http://otabbs.ots.gov:80/E574T101.

I have exerpted parts of Chapter 4: *Alternate Concepts for Fusion Energy Administration*.

### STATUS AND PROSPECTS OF ALTERNATE CONCEPTS

There are several alternate fusion concepts with a wide-range of maturity levels or development of the information base. Over the past decades, the primary focus of the fusion energy program has been on several MFE concepts. (see footnote 9) Extensive research relevant to IFE has also been performed, largely for its potential defense applications. As a result, many MFE and IFE concepts generally enjoy a far more advanced knowledge base than other fusion concepts such as the colliding beam and inertial electrostatic concepts. Past efforts have been much less extensive both in theory and experiment, and knowledge about the prospects is far more speculative.The likelihood that some alternate concept may attain and exceed the expected technical and economic performance of the tokamak remains speculative. Developing comparative information judging the relative strengths and weakness of a broad range of alternate concepts and assessing the information base has not been a priority of the fusion energy program. In particular, there is no current, published DOE-sponsored analysis of the comparative technical prospects and challenges of the broad array of fusion concepts including novel ones or those previously examined and no longer pursued. DOE has sponsored and published, however, reviews of alternate MFE concepts that discuss their relative level of development and likely prospects, (see footnote 10) and has supported some analyses of the relative prospects of IFE. (see footnote 11) The lack of comparative assessment of non-MFE or IFE concepts is consistent with the fusion energy program's primary focus on MFE concepts rather than a broader array of fusion concepts.

. . .

Some alternate MFE concepts previously investigated and found less promising than the tokamak may warrant reconsideration, based on improvements in technology and theoretical understanding. For example, one of the major challenges with the stellarator concept was designing and fabricating the relatively intricate magnets required. However, advanced computer-based analytical capabilities continue to improve the ability to design and manufacture magnets. Some of these techniques were developed and used in producing the now prematurely retired Advanced Toroidal Facility (ATF), the most recent stellarator. (see footnote14) While the stellarator may not ultimately prove more attractive than the tokamak, improving magnet technology continues to reduce one of its principal drawbacks. Advantages relative to the tokamak include that they are inherently steady state, have no plasma current, and thus do not suffer from disruptions and instabilities of the plasma. The approximately $1-billion Large Helical Device (LHD), under construction in Japan, is a superconducting stellarator similar to ATF in concept, but closer to TPX in scope and cost. A similar scale stellarator has been proposed in Germany. A much smaller stellarator with a cost of about $3 million is under construction at the University of Wisconsin as part of DOE's small program for alternate fusion concepts.

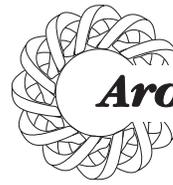

## Around the Labs

## Edge Transport and Turbulence Studies on U-3M

Joint measurements of equilibrium plasma potential, electron temperature and ion density, and the fluctuations in plasma potential and ion density are being carried out in the edge region of the $l = 3$ $m = 9$ Uragan-3M (U-3M) torsatron. In particular, this work is undertaken to clarify the effect of low-frequency electrostatic turbulence on particle and energy transport.

In view of data summarized in Ref. 1, the elucidation of this question under U-3M conditions is of special interest because in U-3M the plasma is heated by RF fields in the multimode Alfvén resonance regime ($\omega \leq \omega_{ci}$) [2]. It follows from both the theory and some experimental work that under these conditions parametric interaction can arise in the edge plasma between the pump oscillations and natural electrostatic ion cyclotron harmonic oscillations; this can result in an increase of low-frequency (drift) turbulence. In this light it would be of interest to estimate the equilibrium particle and heat flows at the boundary of the confinement region and to compare them with the turbulence-induced flows near and far from the RF antenna.

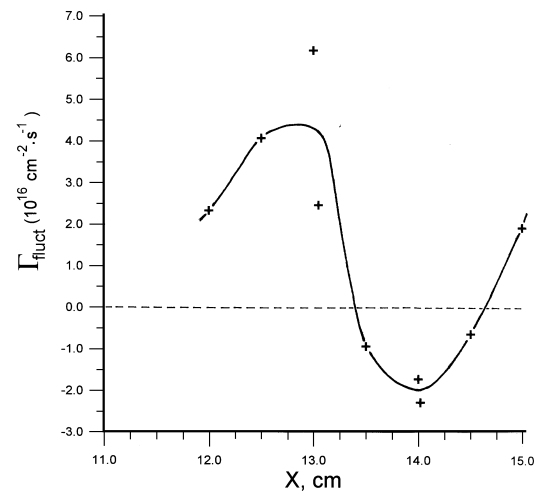

**Fig. 1.** Turbulence-driven particle flow as a function of minor radius (the boundary is at ~ 12.5 cm).

At present the measurements have been made far from the antenna in so-called "low-density regime" ($P_{RF} \approx 200$ kW, $B_\varphi = 0.45$ T, $\bar{n}_e \approx 4\times 10^{12}$ cm$^{-3}$, $T_e(0) \approx 300$ eV, $T_i(0) \approx 150$ eV) [2]. An array of four Langmuir probes is used which is introduced from the low field side and moved along the major radius. In the absence of a material limiter the measurements of floating potential carried out near the calculated last closed magnetic surface (LCMS, $\bar{a} = 12.5$ cm) indicate the radial electric field $E_r$ shear layer, which is ~ 2 cm wide. In this region the radial profiles of electron temperature and ion saturation current are measured. Supposing that there are no charged particle sources in the boundary layer far from the antenna and that the balance takes place between the perpendicular and parallel particle flows in the scrape-off layer (SOL), the estimation of the equilibrium perpendicular diffusion flow through the boundary of the confinement region yields $\Gamma_{eq} \approx 6 \times 10^{16}$ cm$^{-2}$s$^{-1}$. This value is 2 to 3 orders of magnitude larger than that following from neoclassical predictions, but is comparable with $\Gamma_{eq}$ values having been obtained by similar technique in other closed magnetic traps [1] (see $\Gamma_{SOL}$ in Fig. 2).

At the boundary of the confinement region the relative levels of plasma potential (floating potential) and ion density (ion saturation current) fluctuations amount, respectively, to $eV_p/T_e \lesssim 1$ and $\tilde{n}/n \approx = 0.3$.

Similar to the Texas Experimental Tokamak and the ATF torsatron cases, a mean phase velocity shear layer exists in the vicinity of the last closed magnetic surface. The maximum radial particle flow induced by the turbulence attains $\tilde{\Gamma} \approx 5 \times 10^{16}$ cm$^{-2}$s$^{-1}$ as is shown in Fig. 1. The $x$ coordinate is the distance from the center of the poloidal cross-section which is close to the local equilibrium flow. Note that there exists a radial region with $\tilde{\Gamma}$ going inward outside the confinement region. This might be associated with ionization-driven drift wave turbulence [3]; though to confirm or to refute this statement further, more detailed studies are necessary.

In Fig. 2, the result of a comparison of the equilibrium and turbulence-induced particle flows is presented together with the data on similar comparisons made on some other closed magnetic traps [1]. We may conclude that similar to some tokamaks and the ATF torsatron, in U-3M the perpendicular particle flow at the edge is dominated by the turbulence-induced flow. In their absolute values, at least far from the RF antenna, these flows are comparable with the edge flows in other closed magnetic traps. In this sense we may say that the chosen method of plasma heating in U-3M (multimode Alfvén resonance) does not result in any significant deterioration of particle confinement in comparison with other methods of plasma heating (ohmic, electron cyclotron resonance, neutral beam injection).

Victor V. Chechkin for U-3M group
Institute of Plasma Physics
National Science Center
Kharkov Institute of Physics and Technology
Kharkov, Ukraine

FAX:007-057-235-2664
E-mail:ipp@ipp.kharkov.ua

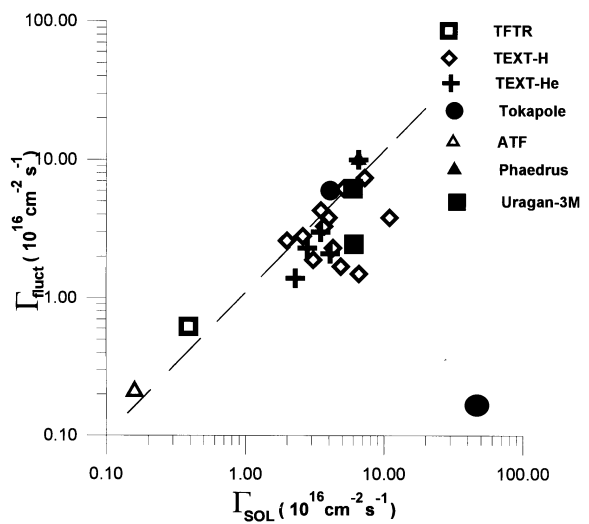

**Fig. 2.** A comparison between turbulence-driven particle flow and equilibrium radial flow from Ref. [1] with the U-3M data added.

# Topological aspects of island divertor studies on W7-AS

W7-AS exhibits a flexible boundary topology that can be controlled by means of the external rotational transform. The structure of the plasma boundary is characterized by the presence of 5/$m$ natural islands, and this topology can be used to study the feasibility of island divertors for power and particle exhaust. The 5/$m$ islands can exist as a closed chain, or they can be intersected by the inboard limiters. These limiters provide a nearly homogeneous scrape-off layer and can be used for preliminary investigations to explore the divertor potential of W7-AS. This is especially important in context of the planned island divertor for W7-X.

The viability of the island divertor concept depends critically on the stability of the island structure against external and internal magnetic field perturbations. To assess the relevance of vacuum field structures and of perturbative fields for plasma transport and recycling at the edge, we have investigated the boundary topology for different values of the edge rotational transform $\iota_a$ within the range of $\iota_a$ between 0.34 and 0.64. The experiments were carried out for net current-free, low-beta (~ 0.1%), flat-top electron cyclotron resonance heating (ECRH) discharges with $P_{heat}$ = 168 kW and $\langle n_e \rangle$ = $3.8 \times 10^{18}$ m$^{-3}$. A Langmuir probe array located at a fixed radius outside the separatrix and covering about one-third of the poloidal plasma circumference measured the poloidal $n_e$ (and $T_e$) profiles at the boundary, while the distribution of the neutral gas density was monitored by measuring the spatial variation of the H$_\alpha$ intensity. Video cameras were used as monitors for the visible radiation at the edge. Experimental results from these diagnostics show clear evidence of the presence of the 5/8, 5/9, 5/10 and 5/11 island chains at the plasma edge. Figure 1 shows a tangential view of the plasma and the upper inboard limiter in Module 5 for a discharge with $\iota_a = 0.525$. The 5/10 island structure at the boundary is clearly visible in the picture.

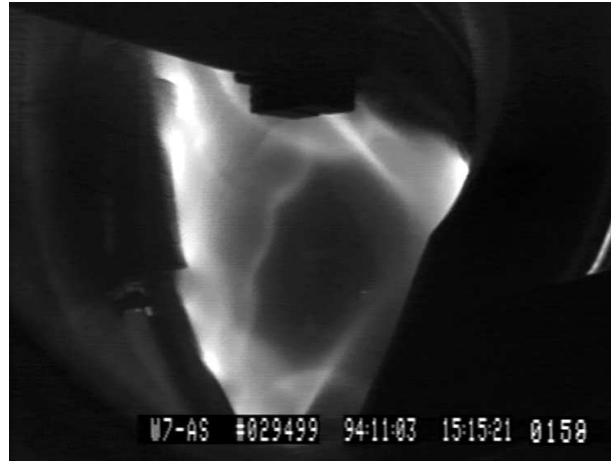

**Fig. 1.** Tangential view of the plasma taken with a video camera installed in Module 5.

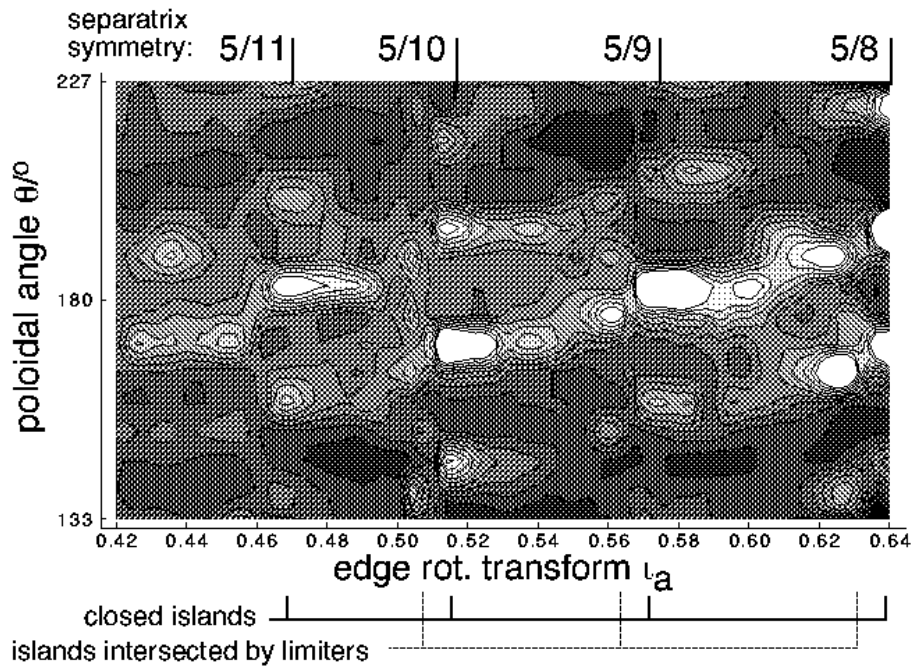

**Fig. 2.** Plot of $n_e$ contours measured with the Langmuir probe array. The darkest areas indicate $n_e$ < 3 x 10$^{17}$ m$^{-3}$, and the brightest areas indicate $n_e$ > 1.5 x 10$^{18}$ m$^{-3}$ in a linear scale.

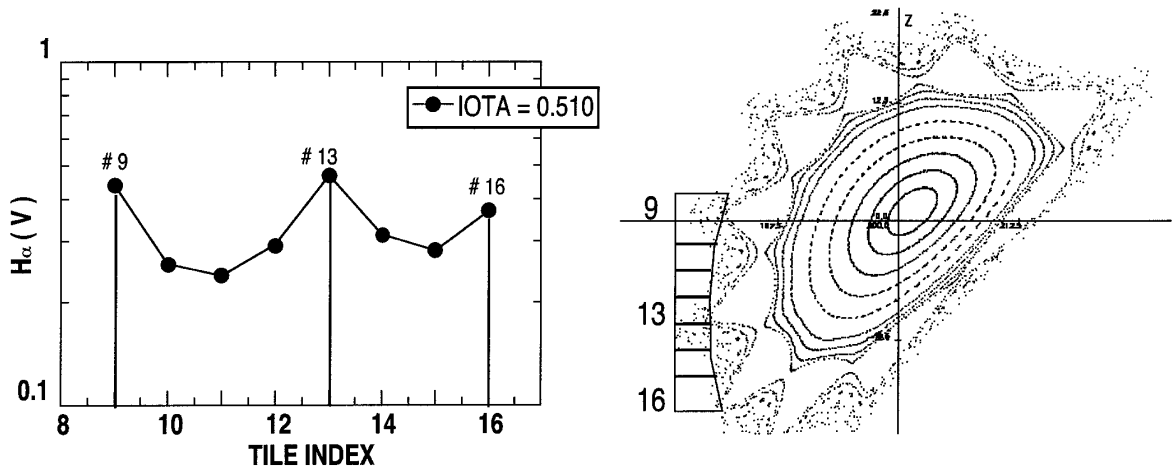

**Fig. 3.** Poloidal profile of the $H_\alpha$ intensity in front of the lower inboard limiter. The vacuum field configuration is also shown.

Figure 2 shows $n_e$ profiles at the edge as a function of $\iota_a$. The higher density, higher temperature areas around the closed islands are indicated by the bright areas in the figure. The change in the separatrix symmetry between odd and even values of $m$ (X-point at the midplane for 5/8 and 5/10 islands, O-point for 5/9 and 5/11 islands) can be clearly seen in the $n_e$ profiles. As $\iota_a$ decreases the islands shift radially outward and are eventually intersected by the inboard limiters. This leads to a loss of confinement within the islands, resulting in the decay of the plasma pressure within the island core. The sharp edges of the bright areas in Fig. 2 arise due to this sudden loss in confinement. The transition from closed to open islands is reflected in the experimental data as a doubling of the maxima in the temperature and density profiles (best resolved for $\iota_a$ = 5/9 and 5/10). This is because the intersection with the limiters results in the formation of two SOL branches, and the interaction with the limiter now occurs along two zones per island. The figure also demonstrates that the 5/$m$ structures are not significantly perturbed by lower order resonances even at $\iota_a = 1/2$, where such perturbations are expected to have a major influence. Measurements carried out at higher beta ($\beta = 0.8\%$) indicate that despite deviations in detail, the 5/$m$ island chains are still dominant under these conditions.

Comparison of the experimentally measured poloidal phase of the islands, in particular their location with respect to the inboard limiters, with that predicted by vacuum field calculations is a good measure of the relevance of these structures for a real plasma. Figure 3 shows a poloidal $H_\alpha$ intensity profile along the surface of the lower inboard limiter in Module 3 along with the corresponding magnetic configuration. The edge rotational transform was fixed at 0.510 for this discharge. It is clearly seen that the predicted positions of the islands (at limiter tiles 9, 13 and 16) are in excellent agreement with the location of the maxima in the $H_\alpha$ profile. This is true also for the 5/8, 5/9, and 5/11 island chains. Poloidally resolved calorimetric measurements on the inboard limiters also confirm these results.

J. Das for the W7-AS Team
Max Planck Institut für Plasmaphysik
IPP Garching, Germany

Phone: 0049-89-3299-1725
FAX:    0049-89-3299-2584
E-mail: jad@ipp-garching.mpg.de

# 140-GHZ second harmonic O-mode electron cyclotron heating in W7-AS

Electron cyclotron resonance heating (ECRH) at the second harmonic ordinary mode polarization (O2-mode) has a low single-pass absorption at the typical W7-AS plasma parameters but gives access to a density twice the cut-off density limit of the second harmonic extraordinary mode (X2-mode). For W7-AS therefore, X2-mode heating with its high single-pass absorption is the standard heating scenario. In the next step stellarator W7-X, a much higher electron temperature and a much higher single-pass absorption of the O2-mode are expected; therefore the O2-mode ECRH is an attractive scenario [1]. With this in mind we have investigated the general absorption properties of the O2-mode in W7-AS.

A 140-GHz, 40-ms pulse with 0.7 MW power in O2-polarization was launched at an optimum oblique angle of 20° into a NBI-sustained target plasma with a magnetic field of 2.5 T. A central electron density of $1.8 \times 10^{20}$ m$^{-3}$ was chosen, which is well above the X2 cut-off density. The electron temperature is only 0.4 keV under these high-density conditions. The plasma energy content represented by the diamagnetic signal in Fig. 1 increased by about 1.5 kJ compared to a similar discharge with NBI only. The absorbed heating power is 110 kW, determined from the time derivative of the diamagnetic signal at the heating pulse switch-off. This is roughly twice the calculated power absorbed in a single pass and can be explained by wave reflection at the inner torus wall after a first plasma transit.

The soft X-ray emission was analyzed to determine the power deposition zone. The line-integrated X-ray emissions along nine different chords more or less in the direction of the plasma elongation are recorded with a spatial resolution of about 2 cm. Despite the low spatial resolution the time delay between the decay of the different X-ray detector signals after heating power switch-off is easily resolved. This time delay with respect to the moment of power switch-off is sketched in Fig. 2 as a function of the radial position of the X-ray emission. The time delay increases with the effective chord radius, indicating that the heating power must have been absorbed in the dense plasma core at the ECR layer.

H. Laqua for the W7-AS Team
Max-Planck-Institut für Plasmaphysik, EURATOM Ass.
D-85748 Garching, FRG

E-mail laqua@ipp-garching.mpg.de
Phone: 0049-89-3299-1963
FAX:   0049-89-3299-2584

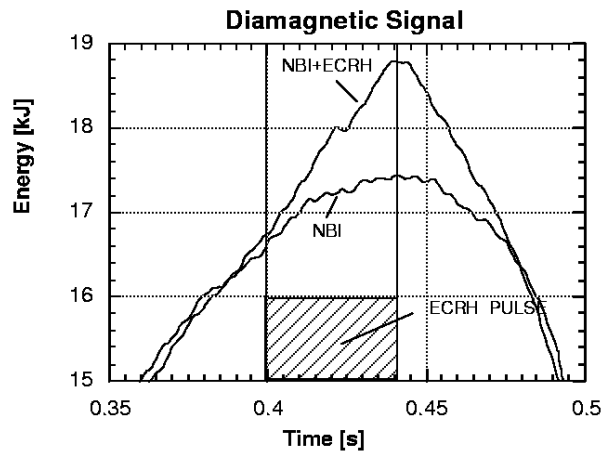

**Fig. 1.** Energy content from the diamagnetic signal of a discharge with combined NBI and O2-mode ECRH (upper curve, suppressed origin). For comparison a purely NBI-heated discharge is shown in the lower curve. The time window of the ECRH pulse is shown by the markers.

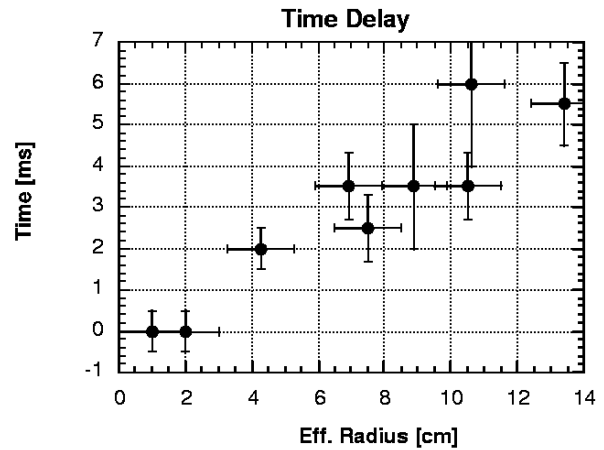

**Fig. 2.** Time delay of the soft X-ray emission decay at different effective chord radii with respect to the ECRH switch-off time.

# Equilibrium plasma currents in quasi-symmetric stellarators with a helical magnetic axis

The concept of a quasi-symmetric stellarator was advanced in [1,2]. A latent symmetry of the three-dimensional (3-D) magnetic field in such systems must improve their confinement properties because of the particle drifts and heat fluxes across the magnetic surfaces. The finite-beta plasma currents can also be reduced. Initially, such configurations were investigated with the use of 3-D equilibrium codes, which can require significant computer resources. For example, the Wendelstein 7-X configuration was found computationally.

More recently, theoretical and numerical methods have been developed to make examination of some of the questions related to quasi-symmetry (QS) easier. Theoretical investigations show [3–5] that in the paraxial approximation the QS conditions are given as relations of following parameters: $k$, $\kappa$, $\varepsilon$ and $\delta$, where $k(s)$ and $\kappa(s)$ are the first and the second curvatures of the magnetic axis, $\varepsilon(s)$ is the ellipticity of the magnetic surface cross-sections, $\delta(s)$ is the angle between the normal to the magnetic axis and the small semiaxis of the ellipse, and $s$ is the arc length of the magnetic axis. The correlation equations for these parameters were obtained (in the first order of the paraxial expansion) in [3].

In this work the equilibrium plasma currents are investigated in the paraxial approximation for quasi-symmetric configurations with a helical magnetic axis where QS conditions are determined by the equations obtained in [3]. The parameter

$$\gamma_{\|}/\gamma_{\perp} = \int j_{\|}^2 \delta V / \int j_{\perp}^2 \delta V, \quad (1)$$

which is one of the important stellarator optimization parameters, is also calculated. In Eq. (1) $j_{\|}$ and $j_{\perp}$ are the plasma current components, parallel and perpendicular to the magnetic field, and the integrals imply integration over the volume of the layer between closely spaced magnetic surfaces.

To solve the equations [3] for the QS conditions, numerical methods are necessary. Because of the conditions the magnetic surface geometry in the systems under consideration is very complicated. Therefore, the $j_{\|}$ and $j_{\perp}$ and the geometric factor [Eq. (1)] are calculated by numerical integration along the field lines using the methods of paper [6].

We start from an assignment of the magnetic axis shape and find the parameters $\varepsilon$ and $\delta$ as functions of $s$ using the equations of paper [3]. After that we obtain the expressions for the magnetic field, $\boldsymbol{B}$, which are necessary for the plasma current differential equations. For this purpose we use the work coordinate system $(a, \theta, s)$,

$$x = a\, e^{-\eta/2} \cos\theta = \rho\, cos(\vartheta+\delta),$$
$$y = a\, e^{\eta/2} \sin\theta = \rho\, sin(\vartheta+\delta), \quad (2)$$

where $\varepsilon(s) = \tanh\eta(s)$; $a$ is the magnetic surface label, $\theta$ is a quasi-poloidal angle, $\rho$ is the distance from the magnetic axis, and $\vartheta$ is the angle measured from the normal to the magnetic axis. The expression for $\boldsymbol{B}$ may be represented as in [3] (on the magnetic axis, the magnetic field, $B_0$, is constant, and we assume that the total plasma current, $J$, is zero):

$$2\pi \boldsymbol{B} = F\nabla_a\zeta + \nabla_a\varphi, \quad (3)$$

or

$$\boldsymbol{B} = \frac{F}{2\pi}\left[\frac{1}{F}\frac{\partial\varphi}{\partial\theta}\nabla_a\theta + (1 + \frac{1}{F}\frac{\partial\varphi}{\partial\zeta})\nabla_a\zeta\right]. \quad (4)$$

Here $\nabla_a = \nabla - \nabla a(\nabla a\nabla)/|\nabla a|^2$, $F = 2\pi R B_0$, $R = L/2\pi$, $L$ is the full length of the magnetic axis, $\zeta = s/R$, $a/R$ is the small parameter of the paraxial expansion,

$$\frac{1}{F}\frac{\partial\varphi}{\partial\theta} = \frac{a^2}{R^2}\left[\varepsilon\left(\frac{d\delta}{d\zeta} - \kappa R\right)\cos 2\theta + \frac{\cosh\eta}{2}\frac{d\eta}{d\zeta}\sin 2\theta\right] \quad (5)$$

In the approximation considered, the term $(\partial\varphi/\partial\zeta)/F$ may be omitted in comparison with unity. The $\nabla a$, $\nabla\theta$ and $\nabla\zeta$ components can be calculated using the expressions in Eq. (2).

In accordance with the calculation methods developed in [6] the currents $j_{\|}$ and $j_{\perp}$ and the parameter [Eq. (1)] can be calculated using the system of ordinary differential equations. It includes first, the equations of the magnetic field lines

$$da/d\tau = 0, \quad d\theta/d\tau = \boldsymbol{B}\nabla\theta, \quad ds/d\tau = \boldsymbol{B}\nabla s. \quad (6)$$

Second the equations for the equilibrium plasma currents are given as

$$\boldsymbol{j}_{\perp} = cp'[\boldsymbol{B}\nabla\Psi]/B^2 \quad \boldsymbol{j}_{\|} = cp'h\boldsymbol{B}, \quad (7)$$

$$dh/d\tau = -2[\boldsymbol{B}\nabla B]\nabla\Psi/B^3, \quad (8)$$

where $c$ is the speed of light, and $h$ arises from the equilibrium equations. Third, the equations for calculating the integrals of these currents over the volume of the layer between neighboring magnetic surfaces follow

$$df_1/d\tau = (hBB_0)^2,$$

$$df_2/d\tau = (B_0 \nabla\Psi/B)^2 \qquad \left(f_1(0) = f_2(0) = 0\right) \qquad (9)$$

(the parameter $\tau$ is an integration variable, the prime denotes the derivative with respect to the magnetic surface function $\Psi$ ($\boldsymbol{B}\nabla\Psi = 0$), the definition $\nabla\Psi = \nabla a/|\nabla a_0|$ is used, $\nabla a_0$ is the $\nabla a$ value at the starting point of the integration). The right-hand sides of Eq. (9) are proportional to the squares of $j_\parallel$ and $j_\perp$. Therefore, the parameter [Eq. (1)] is determined as

$$\gamma_\parallel / \gamma_\perp = f_1(\tau_m)/f_2(\tau_m), \qquad \tau_m \to \infty, \qquad (10)$$

$\tau_m$ is the upper limit of the integration interval.

We considered a magnetic axis specified by the parametric form

$$r = r_0 + r_1 \cos(N\phi), \qquad z = r_1 \sin(N\phi), \qquad (11)$$

which is a helical line on a torus ($r$, $\phi$, $z$ are the cylindrical coordinates, $N$ is a number of periods of the system). The calculations were carried out for several configurations at $r_1/r_0 = 0.132$, $N = 5$ and $N = 3$. The results of the calculations of the factor Eq. (1) (for $a/r_1 = 1/13.2$ at $N = 5$ and $a/r_1 = 1/26.4$ at $N = 3$) are presented in Table 1, where the parameters $E_0$, $n_1$, $\mu$, $\eta$, and $\delta$, which characterize the configurations under consideration, are also presented. Here $E_0 = e^{\eta(0)}$ is the ratio of large and small semiaxes of ellipse at $\phi = 0$, $n_1 = n/N$; $n$ is the number of turns made by the ellipse relative to the normal to the magnetic axis on the full length of the axis; $\mu$ is the rotational transform relative to the normal to the magnetic axis; $\eta_{1,2,3}$ and $\delta_{1,2,3}$ mean the corresponding distributions of $\eta$ and $\delta$ over $s$, which are determined by the assignment of $E_0$ and $\delta(0) = 0$. The parameters in the ($\eta_1$, $\delta_1$) and ($\eta_2$, $\delta_2$) cases correspond to the magnetic configurations considered in Ref. [5].

Some of the results pertain when the QS conditions are violated. The $\delta = \delta_1$, $\eta = \eta_2$ case corresponds to slightly violated conditions, and the $\delta = 0$, $\eta = 0$ cases correspond to strongly violated conditions and to a circular shape of the magnetic surface cross sections.

It follows from the results obtained that for the considered variants of the systems, the geometric factor [Eq. (1)] is less than unity, satisfying one of the important stellarator optimization conditions. For the same magnetic axis parameters but at strong violation of the QS conditions (for the circular shape of the magnetic surface cross sections) the mentioned factor is essentially greater than unity.

The decrease of the parameter [Eq. (1)] corresponds to a lowering of the secondary plasma currents and to a reduction of distortion of the magnetic configuration due to finite plasma pressure. Moreover, in the Pfirsch-Schlüter regime, the geometric factor of the diffusion coefficient increase due to the particle drift in the inhomogeneous magnetic field is determined by the parameter [Eq. (1)]. Thus, using relatively simple methods of investigation, we have demonstrated the possibility of improvements in these parameters for quasi-symmetric systems in the Pfirsch-Schlüter regime.

The results of calculations of the plasma currents were also used for the analysis of the QS condition in the invariant form [5] for the systems under consideration. The analysis has shown that the QS condition in the invariant form is well satisfied at sufficiently small distances from the magnetic axis for the configuration variants, for which the QS conditions were obtained in the paraxial approximation with the help of the equations of paper [3].

V. V. Nemov
Institute of Plasma Physics, National Science Center
Institute of Physics and Technology
Kharkov 310108, Ukraine.

FAX:007-057-235-2664
E-mail:ipp@ipp.kharkov.ua

Table 1. The results of $\gamma_\parallel/\gamma_\perp$ calculations

| N | $E_0$ | $n_1$ | $\delta$ | $\eta$ | $\mu$ | $\gamma_\parallel/\gamma_\perp$ |
|---|---|---|---|---|---|---|
| 5 | 2 | 0 | $\delta_1$ | $\eta_1$ | 3.165 | 0.667 |
| 5 | 1.5 | -1/2 | $\delta_2$ | $\eta_2$ | 3.39 | 0.82 |
| 5 | 1 | – | 0 | 0 | 4.224 | 4.0 |
| 5 | 1.5 | 0 | $\delta_1$ | $\eta_2$ | 3.29 | 0.73 |
| 3 | 1.65 | -1/2 | $\delta_3$ | $\eta_3$ | 1.9 | 0.501 |
| 3 | 1 | – | 0 | 0 | 2.806 | 55.3 |

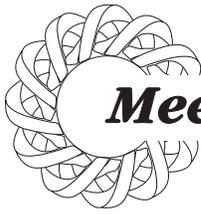

# Meetings

## Tenth International Conference on Stellarators
## Madrid, Spain   May 22 – 26, 1995

**PRELIMINARY ANNOUNCEMENT**

**General Information**

The Tenth International Conference on Stellarators will be a Technical Committee Meeting of the International Energy Agency and will be hosted by the Centro de Investigaciones Energéticas Medioambientales y Tecnológicas (CIEMAT). The conference will be held at the CIEMAT main auditorium in Madrid, Spain.

**Program**

The conference will cover all topics applicable to stellarators and other helical confinement devices but will not be restricted to these devices. Special emphasis will be given to edge physics problems. Some topics of interest are:

- Experimental results
- Theory
- Diagnostics of special interest to helical devices
- Comparison tokamak/stellarator
- New devices, next generation experiments, reactors

The number of oral presentations will be limited to emphasize overview talks, reviews of recent work, and topics of general interest. More detailed papers will be presented in poster sessions. Ample time will be set aside for free discussion and more intense, focused, topical meetings.

The International Program Committee will be composed of

C. Alejaldre, CIEMAT, Madrid, Spain

G. Grieger, Max Planck Institut für Plasmaphysik, Garching, Germany

A. Iiyoshi, NIFS, Nagoya, Japan

L. M. Kovrizhnykh, Institute for General Physics, Moscow, Russia

J. F. Lyon, Oak Ridge National Laboratory, Oak Ridge, USA

Interested persons should notify the Scientific Secretary of their intention to attend the meeting and provide a tentative title of their presentation (and whether oral or poster presentation is preferred) by January 16, 1995, on the enclosed form, to allow inclusion of a tentative agenda in the Final Announcement. A one-page abstract of each paper is needed by April 17, 1995, to allow distribution of a book of abstracts at the start of the Conference. Please send this information to:

Enrique Ascasibar, Scientific Secretary
Fusion Area
CIEMAT
Av. Complutense 22
28040 Madrid, Spain

Telephone: +34-1-346.63.69     FAX: +34-1-346.61.24
E-mail: ascasibar@ciemat.es

**Tenth International Stellarator Conference**
**22-26 May 1995, Madrid, Spain**

**Preliminary Registration Form**

PLEASE CHECK:

___ I plan to attend the conference.

___ I plan to bring a gues.t

___ I plan to submit a contribution.

___ Oral presentation preferred.

___ Poster presentation preferred.

Please print or type:

NAME: _______________________________

Phone/FAX: ___________________________

Affiliation: ____________________________

E-mail: _______________________________

Address: _____________________________

_____________________________________

_____________________________________

Paper title: ___________________________

Mail this form to the Scientific Secretary.